\documentstyle[12pt]{article}
\def\bbbc{{\mathchoice {\setbox0=\hbox{$\displaystyle\rm C$}\hbox{\hbox
to0pt{\kern0.4\wd0\vrule height0.9\ht0\hss}\box0}}
{\setbox0=\hbox{$\textstyle\rm C$}\hbox{\hbox
to0pt{\kern0.4\wd0\vrule height0.9\ht0\hss}\box0}}
{\setbox0=\hbox{$\scriptstyle\rm C$}\hbox{\hbox
to0pt{\kern0.4\wd0\vrule height0.9\ht0\hss}\box0}}
{\setbox0=\hbox{$\scriptscriptstyle\rm C$}\hbox{\hbox
to0pt{\kern0.4\wd0\vrule height0.9\ht0\hss}\box0}}}}
\def\bbbe{{\mathchoice {\setbox0=\hbox{\smalletextfont e}\hbox{\raise
0.1\ht0\hbox to0pt{\kern0.4\wd0\vrule width0.3pt
height0.7\ht0\hss}\box0}}
{\setbox0=\hbox{\smalletextfont e}\hbox{\raise
0.1\ht0\hbox to0pt{\kern0.4\wd0\vrule width0.3pt
height0.7\ht0\hss}\box0}}
{\setbox0=\hbox{\smallescriptfont e}\hbox{\raise
0.1\ht0\hbox to0pt{\kern0.5\wd0\vrule width0.2pt
height0.7\ht0\hss}\box0}}
{\setbox0=\hbox{\smallescriptscriptfont e}\hbox{\raise
0.1\ht0\hbox to0pt{\kern0.4\wd0\vrule width0.2pt
height0.7\ht0\hss}\box0}}}}

\begin{document}
\begin{titlepage}
\rightline {quant-ph/9809047 \  \  \  \   }

\vskip 2truecm

\centerline{\Large\bf  Quantum mechanics of an electron }
\vspace*{1mm}
\centerline{\Large \bf in a homogeneous magnetic field and}
\vspace*{1mm}
\centerline{\Large \bf a singular magnetic flux tube}

\vskip 0.5truecm

%

\centerline{\bf Hans-Peter Thienel}


\centerline{Universit\"at Siegen, Fachbereich Physik,
D-57068 Siegen, Germany}
\centerline{e-mail: thienel@hepth2.physik.uni-siegen.de}

\vskip 0.5truecm

\abstract

\noindent
The eigenvalue problem of the Hamiltonian of an electron
confined to a plane and subjected to a perpendicular
time-independent magnetic field which is the sum of
a homogeneous field and an additional field
contributed by a singular flux  tube, i.~e.\ of zero width,
is investigated.
Since both a direct approach based on distribution-valued
operators and a limit process starting from
a non-singular flux tube, i.~e.\ of finite size,  fail,
an alternative method is applied leading to consistent
results. An essential feature is quantum mechanical
supersymmetry at $g=2$ which imposes, by proper
representation, the correct choice of
``boundary conditions''. The corresponding representation
of the Hilbert space in coordinate space differs
from the usual space of square-integrable 2-spinors,
entailing other unusual properties. The analysis
is extended to $g\ne 2$ so that supersymmetry is
explicitly broken. Finally, the singular Aharonov-Bohm
system with the same amount of singular flux
is analysed by making use of the fact that 
the Hilbert space must be the same.

\end{titlepage}

\section{Introduction}

The parameter space of the system that is considered interpolates between
two very different systems of fundamental importance.

On the one hand, the motion of an electron in a plane perpendicular to
a homogeneous magnetic field \cite{pag}\cite{lan}
has found an important application in quantum Hall physics
\cite{lau}\cite{qhe1}\cite{jai}.
Its most striking feature is the vast degeneracy within the
discrete Landau levels that makes quantum Hall phenomena possible.
It is of interest how this system is altered by inhomogeneities
of the magnetic field. An additional singular flux tube appears
to be a minimal modification of the homogeneous field \cite{lew}, while
in general inhomogeneous magnetic
fields are notoriously difficult to handle.

On the other hand,
since the work of Aharonov and Bohm \cite{aha}
the physics of a magnetic flux tube alone has been investigated under
a variety of aspects. Considering particles of spin $1/2$ \cite{mor},
the system is of interest as a simple application
of mathematical index theorems
\cite{as}\cite{aps}\cite{cas}, interrelated with anomalies in quantum field
theories \cite{jac}\cite{sto}\cite{for} and the fact that the Pauli equation
exhibits a quantum mechanical supersymmetry
for $g=2$ \cite{gen}. Furthermore, it is
particles of spin $1/2$ that problems occur for, when we go over from
the non-singular, i.~e.\ flux tube of finite width, to the singular case,
i.~e.\ flux tube of zero width. While for the non-singular case
the standard quantum mechanics works well, the singular case raises
the question of the correct boundary conditions
at the location of the flux tube.
Standard boundary conditions, e.~g.\ of the usual Dirichlet type,
are not sufficient to characterize
the behavior of the eigenfunctions of the Hamiltonian \cite{alf}\cite{sou}.
This is due to the fact that functions of
different boundary behavior are mapped into each other by the
supersymmetry, which is thus not represented automatically
by respecting simple boundary conditions. This difficulty makes
one suspect that the transition from the non-singular to the
singular flux tube might not be continuous.
This point among other questions is obscured by the fact that
a flux tube alone, yielding the free particle behavior as we go to infinity,
leads to a continuum in the spectrum of the Hamiltonian
entailing non-normalizable eigenfunctions.

We supply the singular flux tube of flux $\alpha \Phi$,
where $\alpha$ is real and $\Phi=hc/|e|$, with an additional
homogeneous magnetic field and solve the eigenvalue problem of the 
Hamiltonian.
Technically our procedure deviates from standard methods considerably.
Some features that we are familiar with from other
quantum mechanical systems are not found, while other features that are
unusual in a quantum mechanical context do occur.
Therefore, we want to clarify some issues relevant to our system and
point out some peculiarities of the system in advance.

The integral part of $\alpha$
is not a gauge effect.
The similarity to a gauge transformation merely shows
that an inhomogeneity constrained to a single point leaves many features
of the  system intact. Yet, systems differing by integral
quanta  of singular flux are different.

We solve the non-relativistic Pauli equation and not
the corresponding Dirac equation.
A relativistic treatment would introduce the Compton wave length $\lambda_C$
as an additional scale to the system. An infinitely thin flux tube
would run into trouble with the consistency of
one-particle quantum mechanics
by introducing a scale falling below $\lambda_C$.
While a magnetic field does not accelerate
a charged particle, a gradient exerts a force on the magnetic dipole.
We could be confident about the one-particle description if
the radius of the flux tube were at least of the magnitude
$|\alpha|^{1/2} \lambda_C$. Otherwise, pair production
due to the inhomogeneity of the
magnetic field had to be taken into account.  
Staying non-relativistic we avoid any conflicts.
Clearly, this means that a singular flux tube 
idealizes a real flux tube of minimal radius
$ |\alpha|^{1/2} \lambda_C$.

The quantum system will not be realized in coordinate
space before the limit $R \to 0$ is taken, where
$R$ denotes the radius of the tube to which the flux $\alpha \Phi$
is confined. 
In the regularized coordinate representation
we quantize the system and calculate matrix elements.
The regularization is an intrinsic necessity in order to handle
unavoidable products of singular objects at the flux tube 
consistently, since matrix elements contain products of both singular
operators and singular wave functions.
Supersymmetric pairing turns out to be possible for eigenfunctions only
due to an arbitrariness following from the regularization.
Also, this regularization makes
possible a deviation of the Hilbert space from
the space of square-integrable 2-spinors $L_2\otimes \bbbc^2$
for $|\alpha|\ge 1$. In particular,
singular supersinglet states occur with properties  similar to those of
a classical point particle in two dimensions.

Usually, after  solving an eigenvalue problem of a Hamiltonian we
verify that the set of eigenfunctions supplies a resolution
of unity in the space of square-integrable functions \cite{neu}.
For our singular system, however,  we do not know a priori what the
Hilbert space should be, i.~e.\ what space of functions
we should find a resolution of unity for. Conversely, we use the orthonormal
set of eigenfunctions of the Hamiltonian to span the
Hilbert space. The completeness of the eigenfunctions of the Hamiltonian
within the Hilbert space is an indispensable requirement that is accounted
for in this way by the very construction.

As opposed to non-singular systems, in the presence of a singular flux tube
the representation of supersymmetry is not automatic by a standard treatment
using local boundary conditions \cite{sto}.
Instead, insisting on a preservation of supersymmetry 
uniquely determines the correct eigenfunctions of the Hamiltonian
and, hence, the correct Hilbert space.

We specifically treat the electron with charge $-|e|$.
The considerations apply to any 
elementary, charged, massive particle of spin $1/2$ by obvious modifications.

\section{Supersymmetry}

The following statements \cite{nic}\cite{wit}\cite{gen}
rely on the assumption that all quantities are
sufficiently differentiable and non-singular so that 
their products are defined.

We consider the Pauli Hamiltonian in the $x$-$y$-plane with $p_z=0$, $g=2$,
and a perpendicular magnetic field $\vec{B}=B_z(\vec{r})\vec{e}_z$,
where $\vec{r}=(x,y)$
\begin{equation}
H={M \over 2} (v_x^2+v_y^2) + {|e|\hbar \over Mc} B_z S_z,
\label{1}
\end{equation}
with the velocities
$v_x=(p_x+ {|e| \over c}A_x)/M$
and
$v_y=(p_y+ {|e| \over c}A_y)/M$.
Angular momenta are always given in units of $\hbar$.
We define the superoperators
$ Q=S_+ V $
and
$ Q^{\dagger}=S_- V^{\dagger} $ with
$V=(M / 2)^{1/2}  (v_y+iv_x)$,
$V^{\dagger}=(M /2)^{1/2}(v_y-i v_x)$
and $S_{\pm}:=S_x \pm  i S_y$.

The Hamiltonian and the superoperators
obey the supersymmetry algebra
$$
Q^2=(Q^{\dagger})^2=0, \quad
H = Q Q^{\dagger} + Q^{\dagger} Q.
$$
\begin{equation}
\Rightarrow [H,Q]=[H,Q^{\dagger}]=0.
\label{2}
\end{equation}
In addition we have
\begin{equation}
[S_z,Q]=Q,\quad [S_z,Q^{\dagger}]=-Q^{\dagger}, \quad [S_z,H]=0.
\label{2a}
\end{equation}
Hermitian superoperators are
$Q^{(1)}=Q+Q^{\dagger}=(2M)^{1/2}(S_x v_y -S_y v_x)$ and
$Q^{(2)}=i(Q^{\dagger}-Q)=(2M)^{1/2}(S_x v_x + S_y v_y)$.

We can immediately draw important conclusions from supersymmetry.
On the one hand, the measured energy is
\begin{equation}
\langle \Psi | H| \Psi \rangle= \langle Q\Psi |  Q\Psi \rangle +
\langle Q^{\dagger}\Psi | Q^{\dagger} \Psi \rangle \ge 0.
\label{3}
\end{equation}
On the other hand, any eigenstate of the Hamiltonian with eigenvalue $E>0$ is
doubly degenerate. By (\ref{2a})
we can always arrange the partner states to be eigenstates of $S_z$ such that
$$
Q |E,\sigma=-1/2 \rangle \sim |E,\sigma=+1/2 \rangle,
$$
\begin{equation}
Q^{\dagger}|E,\sigma=+1/2 \rangle \sim |E,\sigma=-1/2 \rangle,
\label{4}
\end{equation}
where
$S_z|\sigma \rangle =\sigma| \sigma \rangle$ with $\sigma=\pm 1/2$,
if $E>0$.
The eigenstates of $H$ with $E=0$, being annihilated by the superoperators
\begin{equation}
Q|E=0 \rangle  =Q^{\dagger} |E=0 \rangle =0,
\label{5}
\end{equation}
are supersinglets.

\section{Insufficient approaches to the problem}

We give a brief sketch of two obvious
approaches in order to work out their shortcomings resolved by the
correct method in the next section.

\subsection{The direct approach}

We consider the magnetic field in the form
$\vec{B}(\vec{r})=(B+ \alpha\Phi \delta^2(\vec{r})) \vec{e}_z$ \cite{lew}.
$B>0$ is constant everywhere. The corresponding vector potential is
$\vec{A}(\vec{r})=(Br/2+
\alpha\Phi / 2 \pi r)\vec{e}_{\varphi}$,
where $r$ and $\varphi$ denote polar coordinates
in the $x$-$y$-plane; 
i.~e.\ the flux tube part is
a pure gauge  for $r\ne 0$ locally.
We measure lengths in
terms of $\lambda=(\Phi/\pi B)^{1/2}$ putting $\tilde{r}=r /\lambda$.
Energy is measured in units of $\hbar \omega$
with  the Larmor frequency $\omega=|e|B /Mc$ of the homogeneous field.
The superoperators are measured in units of $(\hbar \omega)^{1/2}$.
For $\alpha = 0$ we obtain the operators
for the homogeneous magnetic field.

The Hamiltonian is
\begin{equation}
H_{\alpha}=-{1 \over 4} \left({1 \over \tilde{r}}{\partial
\over \partial \tilde{r}}
\tilde{r} {\partial \over \partial \tilde{r}}+
{1 \over \tilde{r}^2} {\partial^2 \over \partial
\varphi^2}\right) +{1 \over 4} \tilde{r}^2 -
{i \over 2}{\partial \over \partial \varphi}+S_z
+{\alpha \over 2} -{i\alpha \over 2 \tilde{r}^2}
{\partial \over \partial \varphi} +
{\alpha^2 \over 4\tilde{r}^2}+{\alpha\over 2\tilde{r}} \delta(\tilde{r})S_z.
\label{6}
\end{equation}
Since $[H_{\alpha},L_z]=[H_{\alpha},S_z]=0$,
we put
$\Psi_{E,\sigma,m}(\tilde{r},\varphi)= \psi_{E,\sigma,m}(\tilde{r})
e^{im\varphi}\zeta_{\sigma}$
in the eigenvalue equation,
where $m$ is integer and $\zeta_{\sigma}$ are unit eigenspinors of $S_z$.
For $\tilde{r}\ne 0$ the eigenvalue equation is the same as for a
homogeneous magnetic field by the substitution $m \to m+\alpha$.
From the eigenfunctions of the homogeneous field $\alpha =0$ \cite{pag},
we know the solutions for $\alpha \ne 0$
in the entire plane except at $\tilde{r}=0$.
The delta function allows only eigenfunctions
to be continued to $\tilde{r}=0$ that vanish there.
Accordingly, the solutions are
$$
\Psi_{E,\sigma,m}(\tilde{r},\varphi)={1 \over \lambda \sqrt{\pi}}
\sqrt{(E-\sigma- 1/2 -(m+\alpha)\theta(m+\alpha))!
\over \Gamma(E-\sigma+ 1/2 -(m+\alpha)\theta(-m-\alpha))}
$$
\begin{equation}
\times \tilde{r}^{|m+\alpha|} L^{|m+\alpha|}_{E-\sigma-1/2
-(m+\alpha)\theta(m+\alpha)}(\tilde{r}^2)
e^{-\tilde{r}^2\over 2} e^{im\varphi}\zeta_{\sigma}
\label{7}
\end{equation}
with $E-\sigma -1/2-(m+\alpha)\theta(m+\alpha)=0,1,2,...$
and $|m+\alpha|\ne 0$.
For consistency with supersymmetry it is necessary that
each eigenfunction with $E>0$ find a superpartner. The superoperators
\begin{equation}
Q_{\alpha}=S_+V_{\alpha}
=S_+{e^{-i\varphi} \over 2}\left({\partial \over \partial \tilde{r} }
-{i\over \tilde{r}}
{\partial \over \partial \varphi } +\tilde{r}+{\alpha \over \tilde{r}}\right)
\label{8} \end{equation}
and
\begin{equation}
Q_{\alpha}^{\dagger}=S_- V^{\dagger}_{\alpha}=
S_-{e^{i\varphi} \over 2}\left(-{\partial \over \partial \tilde{r} }
-{i\over \tilde{r}}
{\partial \over \partial \varphi } +\tilde{r}+{\alpha\over \tilde{r}}\right),
\label{9} \end{equation}
however, raise or lower the leading power
for small $\tilde{r}$ by one unit such that an
eigenfunction (\ref{7})  that is $O(\tilde{r}^{u})$ with $0<u \le 1$
might be mapped to a superpartner that is $O(\tilde{r}^{u-1})$,
which is not an admissible eigenfunction of $H_{\alpha}$,
being in conflict with the delta function.
Insisting on supersymmetry,  we have to exclude all
eigenfunctions in (\ref{7}) of which the superpartner
is not contained among the eigenfunctions in (\ref{7}) as well.

For each spin component for non-integer $\alpha $
one value of $(m+\sigma)$ is missing.
For integer $\alpha \ne 0$ even two values of $(m+\sigma)$ are missing.
The set of eigenfunctions of the Hamiltonian certainly does not
provide a complete basis for the space of square-integrable 2-spinors 
$L_2\otimes \bbbc^2$ if $\alpha \ne 0$,
because of the missing $m$-values.
Although it appears strange at first sight,
this is no reason to reject this approach.
One should also recall that the
canonical orbital angular momentum $L_z$ is not a
gauge-covariant operator and, therefore, is not observable.
The eigenstates of this direct approach are
contained among the correct eigenstates displayed in
figure 1. However, for $\alpha >0$
the column of states left to the defect line and for $\alpha <0$
the column
right to the defect line is missing. Also the singular states given in
the figure are not visible by the direct approach.
Incidentally, if there were no pairing,
this direct approach would still
yield a vacancy for $\alpha=\pm 1,\pm 2,...$ .

It is not obvious that $\{ Q_{\alpha}, Q_{\alpha}^{\dagger}\}=H_{\alpha}$,
i.~e.\ that the Hamiltonian with
the delta function as in (\ref{6})
is indeed a result of the anti-commutator. This result can be obtained
using complex coordinates $Z=\tilde{r}e^{i\varphi}$, defined also for
$\tilde{r}=0$ by using the identity
\begin{equation}
{\partial \over \partial Z}{1 \over  Z^*}= \lambda^2 \pi \delta^2(\vec{r}),
\label{9a}
\end{equation}
 valid on functions that are continuous at $\tilde{r}=0$ \cite{gel}.
The identity is derived by using a regularization that cannot be avoided.
In general, products 
of singular operators such as
$Q_{\alpha},Q_{\alpha}^{\dagger}$, and $H_{\alpha}$
require a regularization.

Distribution-valued operators require
sufficiently well-behaved wave functions playing the role of test-functions
so that matrix elements are always defined.
This is, implicitly,
the idea behind the above approach to the eigenvalue problem of $H_{\alpha}$,
where the possibility that the
wave functions themselves could exhibit singular behavior
is excluded by inconsistency with the formulation of the problem
in terms of distribution-valued operators.
An extension beyond this setting requires
a regularization for both the operators and the wave functions in
order to make the corresponding products defined.
A treatment of the eigenvalue problem of $H_{\alpha}$
based on such a regularization will have to reproduce
all eigenfunctions of the above method, which are certainly correct.
It could still, however, reveal other more singular eigenfunctions.

\subsection{The singular flux tube as a
limiting case of a flux tube of finite size}

Now we consider the system of a cylindrical flux tube of finite radius $R$.
Apart from the homogeneous field throughout the plane, we add another
homogeneous field for $r\le R$ contributing the flux
$\alpha \Phi $. The magnetic field is
\begin{equation}
\vec{B}(\vec{r})=[ B +
\theta(R-r) \alpha \Phi /\pi R^2 ] \vec{e}_z,
\label{10}
\end{equation}
where $\theta(R-r)=1$ for $r \le R$ and $0$ otherwise.
The corresponding  vector potential is
\begin{equation}
\vec{A}(\vec{r})=\left[{B r \over 2}+
\theta(r-R){\alpha \Phi \over 2 \pi r}
+\theta(R-r){\alpha \Phi r\over 2\pi R^2} \right] \vec{e}_{\varphi}.
\label{11}
\end{equation}
We employ $\tilde{r}=r/\lambda$ and $\tilde{R}= R/\lambda$
with the magnetic length $\lambda$ of the
homogeneous field outside of the flux tube and the
corresponding $\hbar \omega$ as the unit of energy. 
The Hamiltonian is given by
$$
H_{\alpha}=-{1\over 4}\left({1 \over \tilde{r}}
{\partial \over \partial \tilde{r}}
\tilde{r} {\partial \over \partial \tilde{r}}+
{1 \over \tilde{r}^2} {\partial^2 \over \partial
\varphi^2}\right)+{1 \over 4}\tilde{r}^2 -
{i \over 2}{\partial \over \partial\varphi}+S_z
$$
\begin{equation}
+\theta(\tilde{r}-\tilde{R})\left({\alpha \over 2} -
{i\alpha \over 2 \tilde{r}^2}{\partial \over \partial \varphi}+
{\alpha^2 \over 4\tilde{r}^2}\right)
+\theta(\tilde{R}-\tilde{r}){\alpha \over \tilde{R}^2}\left(
{\alpha\over 4\tilde{R}^2}\tilde{r}^2 + {1 \over 2} \tilde{r}^2 -
{i \over 2}{\partial \over \partial \varphi}+S_z \right)
\label{12}
\end{equation}
and the superoperators are given by
$$
Q_{\alpha}=S_+V_{\alpha}= S_+{e^{-i\varphi} \over 2}
\left[{\partial \over \partial \tilde{r} }-{i\over \tilde{r}}
{\partial \over \partial \varphi } +\tilde{r}+{\alpha \over \tilde{r}}
+\alpha  \theta(\tilde{R}-\tilde{r})\left({\tilde{r} \over \tilde{R}^2}
-{1 \over \tilde{r}}\right)\right],
$$
\begin{equation}
Q_{\alpha}^{\dagger}=S_-V_{\alpha}^{\dagger}
=S_-{e^{i\varphi} \over 2}\left[-
{\partial \over \partial \tilde{r} }
-{i\over \tilde{r}}{\partial \over \partial \varphi }
+\tilde{r}+{\alpha \over \tilde{r}}
+\alpha\theta(\tilde{R}-\tilde{r})\left({\tilde{r} \over \tilde{R}^2}
-{1 \over \tilde{r}}\right)\right].
\label{13}
\end{equation}
The supersymmetry algebra (\ref{2}) is fulfilled.
By $[H_{\alpha},S_z]=[H_{\alpha},L_z]=0$ we put
$\Psi_{E,\sigma,m}(\tilde{r},\varphi)=\psi_{E,\sigma,m}
(\tilde{r}) e^{im\varphi} \zeta_{\sigma} $ for the eigenfunction.
The eigenvalue equation of $H_{\alpha}$ 
supplies differential equations of Kummer type
both inside and outside of the flux tube.
The inside solution has to be regular at $\tilde{r}=0$,
while the outside solution must decay as $\tilde{r} \to
\infty$. We match the two solutions
by demanding continuity at $\tilde{r}=\tilde{R}$ yielding
$$
\Psi_{E,\sigma,m}(\tilde{r},\varphi) \sim 
\left[\theta(\tilde{r}-\tilde{R}) \tilde{r}^{m+\alpha}
e^{-\tilde{r}^2\over 2}
U(m+\alpha+\sigma+1/2-E,m+\alpha+1,\tilde{r}^2)
\vphantom{a\over \tilde{R}^2}\right. 
$$
$$
+\theta(\tilde{R}-\tilde{r}) {\tilde{R}^{m+\alpha-|m|}
U(m+\alpha+\sigma+1/2-E,m+\alpha+1,\tilde{R}^2)
e^{\alpha\over 2} \over
{}_1F_1(m\theta(m)+\sigma+1/2-
{\tilde{R}^2 E \over \tilde{R}^2+\alpha},1+|m|,\tilde{R}^2+\alpha)}
$$
\begin{equation}\left.
\times e^{-(1+\alpha/\tilde{R}^2)\tilde{r}^2/2} \tilde{r}^{|m|}
{}_1F_1(m\theta(m)+\sigma+1/2-{\tilde{R}^2 E \over
\tilde{R}^2+\alpha},1+|m|,\left(
1+{\alpha\over \tilde{R}^2}\right)\tilde{r}^2)\right]
e^{im\varphi} \zeta_{\sigma}
\label{14}
\end{equation}
with the Tricomi function $U$ \cite{abr}.
Still we have to demand continuity of the first derivative
at $\tilde{r}=\tilde{R}$. This entails the
following equation determining the eigenvalue $E$ for given values
of $\sigma,m,\alpha$, and $\tilde{R}$.
$$
\tilde{R}^2 {U'(m+\alpha+\sigma+1/2-E,m+\alpha+1,\tilde{R}^2)\over
U(m+\alpha+\sigma+1/2-E,m+\alpha+1,\tilde{R}^2)}
$$
\begin{equation}
-(\tilde{R}^2+\alpha){{}_1F_1{}'(m\theta(m)+
\sigma+1/2-{\tilde{R}^2 E \over \tilde{R}^2+\alpha},
1+|m|,\tilde{R}^2+\alpha) \over
{}_1F_1(m\theta(m)+\sigma+1/2-
{\tilde{R}^2 E \over \tilde{R}^2+\alpha},1+|m|,\tilde{R}^2+\alpha)}
+m\theta(-m)+\alpha=0.
\label{16}
\end{equation}
The prime indicates differentiation with respect to the third argument.

Since we are interested in the limit $\tilde{R}\to 0$ we consider
(\ref{16}) for small $\tilde{R}$ and find the condition
$$
(\pm m \pm \alpha +\alpha -|m|){}_1F_1
(m\theta(m)+\sigma+1/2,1+|m|,\alpha)
$$
\begin{equation}
-2\alpha {}_1F_1{}'(m\theta(m)+\sigma+1/2,1+|m|,\alpha)=0
\label{17}
\end{equation}
up to terms that vanish as $\tilde{R}\to 0$
if $\tilde{R}^{m+\alpha}
U(m+\alpha+\sigma+1/2-E,m+\alpha+1,\tilde{R}^2)=
O(\tilde{R}^{\pm(m+\alpha)})$.
First of all, we observe that (\ref{17}) is independent of $E$.  But
considering the various cases yields a further simple result.
For $\sigma=-1/2$ and $m\le 0$ on the one hand,
and $\sigma=+1/2$ and $m\ge 0$ on the other
hand, (\ref{17}) is identically fulfilled. Otherwise (\ref{17})  is
never fulfilled. Consequently, if one superpartner is in agreement with
(\ref{17}), the other  necessarily violates (\ref{17}).
Thus, supersymmetric pairs
are not admitted and only supersinglets can occur.
Therefore, only eigenfunctions
with $E=0$ are possible. They have the particularly simple form
\begin{equation}
\Psi_{E=0,\sigma=-1/2,m}(\tilde{r})\sim
[\theta(\tilde{r}-\tilde{R})
\tilde{r}^{-(m+\alpha)}e^{-\tilde{r}^2\over 2}
+\theta(\tilde{R}-\tilde{r})\tilde{R}^{-\alpha}  e^{\alpha\over 2}
e^{-(1+\alpha/\tilde{R}^2)\tilde{r}^2/2} \tilde{r}^{-m}]e^{im\varphi}
\zeta_{-1/2},
\label{18}
\end{equation}
for all $\tilde{R}$ and $\alpha $ with  $m=0,-1,-2,...$
They can also be found by directly solving
$Q_{\alpha} \Psi_{E=0}(\tilde{r},\varphi)=Q_{\alpha}^{\dagger}
\Psi_{E=0}(\tilde{r},\varphi)=0$.
These eigenfunctions are the only ones
that survive the limit $\tilde{R} \to 0$.
Since they can always be normalized so that their norm is finite in the
limit $\tilde{R} \to 0$, all of them have to be considered as correct.

For $\alpha > 0$ these singlets have quantum numbers for which
no eigenfunctions are found by  the direct approach.
By contrast, this approach has not provided the
majority of eigenfunctions found to be correct by the direct approach.
Thus, there is a mutual incompatibility of the two approaches of this
section. The correct method 
will have to reproduce the safe results of both alternatives.

We point out that the failure of the
approach of a shrinking flux tube is not a result of its
rough implementation.
If the magnetic field profile is made smooth
by a modification within the annulus between $R-\eta$ and $R$,
the limit $R\to 0$
will lead to the same results.

\section{A consistent method}

\subsection{The eigenvalue problem of the Hamiltonian}

The direct approach in  3.1 suggests that a regularization
should be applied. However, the most obvious way to regularize the
problem, by regarding the system of the singular flux tube as the
limiting system of a sequence of systems of shrinking
flux tubes in 3.2, fails. Although each member
of the sequence is a consistent
quantum mechanical system, a limiting system does not exist in agreement
with the obviously correct solutions of the direct approach.
Therefore, the idea is to replace the sequence of physical systems at finite
$\tilde{R}$ by a sequence of unphysical systems at finite $\tilde{R}$.
Only the limiting system is physical. In other words,
for finite $\tilde{R}$ the mathematical structures cannot
be interpreted in a quantum mechanical sense.  A consistent quantum
mechanical interpretation is possible only in the limit $\tilde{R}\to 0 $.
This is comparable to the use of the dimensional regularization in
quantum field theory, where the system may  not be interpreted in
physical terms before the regulator is removed.

Before establishing the formal framework, we anticipate the
essential points that suffice for a solution of the eigenvalue problem.
We use the operators of the finite flux tube as given
in 3.2 as their regularized form.
Any matrix element of an operator $\Omega$ is to be calculated as
\begin{equation}
\langle \Xi | \Omega| \Psi\rangle =\lim_{R\to 0} \int d^2r \,
\Xi^*(\vec{r})\Omega \Psi(\vec{r})
\label{19}
\end{equation}
where the regularized forms defined below are to be employed
for  $\Xi(\vec{r})$ and $\Psi(\vec{r})$ as well as for $\Omega$.

We solve the eigenvalue problem of the Hamiltonian
\begin{equation}
\langle E',\sigma',m' | H_{\alpha}-E| E,\sigma,m \rangle =0.
\label{20}
\end{equation}
The solutions have to be normalizable, so we impose
\begin{equation}
\lim_{R\to 0} \int d^2r \,
\Psi^*_{E,\sigma,m}(\vec{r}) \Psi_{E,\sigma,m}(\vec{r}) =1.
\label{21}
\end{equation}
For a solution of (\ref{20}) we use the continuous ansatz (\ref{14})
as a regularized form.
Putting this in (\ref{20})
we get a contribution only from the discontinuity of
the first derivative at $\tilde{r}=\tilde{R}$,
which has to vanish for $\tilde{R} \to 0$, such that
\begin{equation}
\lim_{\tilde{R}\to 0} \,\left[ \left. \psi_{E',\sigma,m}(\tilde{R})\tilde{R}
{\partial \over \partial \tilde{r}}
\psi^{\mbox{\scriptsize in}}_{E,\sigma,m}(\tilde{r})\right|_{\tilde{r}
=\tilde{R}}-\psi_{E',\sigma,m}(\tilde{R})\tilde{R}\left.
{\partial \over \partial \tilde{r}}
\psi^{\mbox{\scriptsize out}}_{E,\sigma,m}(\tilde{r})
\right|_{\tilde{r}=\tilde{R}}\right] =0,
\label{22}
\end{equation}
ensuring hermiticity of $H_{\alpha}$ for $\tilde{R}\to 0$.
This condition is, of course, fulfilled by demanding differentiability
for all finite $\tilde{R}$,
which is what the second approach in section 3 was
based on. Allowing, however, the expression
in the square brackets in (\ref{22})
to be finite for any $\tilde{R}>0$ and only
demanding that it vanish in the limit $\tilde{R}\to 0$
is a weaker condition on the eigenfunctions.

A detailed investigation of (\ref{22})  using
the ansatz (\ref{14}) may be summarized as follows.
There are essentially two ways for an
eigenfunction to fulfill (\ref{22}). The first is that the leading power in
$\tilde{R}$ is great enough to make both terms within the square
brackets in (\ref{22})
vanish separately as $\tilde{R} \to 0$.
The second is that leading non-vanishing contributions
cancel each other, while higher order terms still vanish separately. This
leads again to the condition (\ref{17}) to be fulfilled.
Since it cannot be fulfilled for both partners of
a supersymmetric pair simultaneously, it can be fulfilled either for a
supersinglet, i.~e.\ an eigenfunction for $E=0$, or for a partner of  an
eigenfunction that fulfills (\ref{22}) in the first way.

Besides normalizability and (\ref{22}) we require 
$E\ge 0$ and supersymmetric pairing of the eigenfunctions for $E>0$.

The pairing implies that an eigenfunction for $E>0$ fulfilling (\ref{22})
is admissible only if
its partner also fulfills (\ref{22}). But we have to take into account that
the superoperators contain a derivative in $\tilde{r}$
and will in general map a continuous function
with a discontinuous first derivative at $\tilde{R}$ to a function that is
discontinuous at $\tilde{R}$. It is, however, always
possible to restore continuity by adding a contribution
that does not change the norm, i.~e.\ a representative of the null vector.
Moreover, the behavior for $\tilde{r}<\tilde{R}$ can always be arranged
to have the eigenfunction in the form (\ref{14}), which we will
call the basic form of the eigenfunction. 
The freedom of representing the null vector is an important feature
of the regularization and will be discussed
in more detail in the next section.

The normalized eigenfunctions of Laguerre type are
$$
\Psi_{E,\sigma,m}(\tilde{r},\varphi)= {1 \over \lambda \sqrt{\pi}}
\sqrt{(E-\sigma-1/2-m-\alpha )! \over \Gamma(E-\sigma+1/2)}
\left[\theta(\tilde{r}-\tilde{R})
\tilde{r}^{m+\alpha}e^{-\tilde{r}^2\over 2}
L^{(m+\alpha)}_{E-\sigma-1/2-m-\alpha}(\tilde{r}^2)
\vphantom{\tilde{R}^2 E \over \tilde{R}^2} \right. $$
$$
+\theta(\tilde{R}-\tilde{r}) {\tilde{R}^{m+\alpha-|m|}e^{a\over 2}
L^{(m+\alpha)}_{E-\sigma-1/2-m-\alpha}(\tilde{R}^2)\over
{}_1F_1(m\theta(m)+\sigma+1/2-{\tilde{R}^2 E \over
\tilde{R}^2+\alpha},1+|m|,\tilde{R}^2+\alpha)}
$$
\begin{equation}\left.
\times e^{-(1+\alpha/\tilde{R}^2)\tilde{r}^2/2} \tilde{r}^{|m|}
{}_1F_1(m\theta(m)+\sigma+1/2-{\tilde{R}^2 E \over
\tilde{R}^2+\alpha},1+|m|,
\left(1+{\alpha \over \tilde{R}^2}\right)\tilde{r}^2)\right]
e^{im\varphi}\zeta_{\sigma}
\label{23}
\end{equation}
for $E-\sigma-1/2-m-\alpha=0,1,2,...$
with  $m+\alpha>0$ for $\sigma=-1/2$ and  $m<0$, and 
$m+\alpha>-1$ for $\sigma=+1/2$ and $m\ge 0$.
$$
\Psi_{E,\sigma,m}(\tilde{r},\varphi)= {1 \over \lambda \sqrt{\pi}}
\sqrt{(E-\sigma-1/2)!\over \Gamma(E-\sigma+1/2-m-\alpha)}
\left[\theta(\tilde{r}-\tilde{R})
\tilde{r}^{-m-\alpha}e^{-\tilde{r}^2\over 2}
L^{(-m-\alpha)}_{E-\sigma-1/2}(\tilde{r}^2)
\vphantom{\tilde{R}^2 E \over \tilde{R}^2} \right.
$$
$$
+\theta(\tilde{R}-\tilde{r}) {\tilde{R}^{-m-\alpha-|m|} e^{a\over 2}
L^{(-m-\alpha)}_{E-\sigma-1/2}(\tilde{R}^2) \over
{}_1F_1(m\theta(m)+\sigma+1/2-{\tilde{R}^2 E \over \tilde{R}^2
+\alpha},1+|m|,\tilde{R}^2+\alpha)}
$$
\begin{equation}
\left.
\times e^{-(1+\alpha/\tilde{R}^2)\tilde{r}^2/2} \tilde{r}^{|m|}
{}_1F_1(m\theta(m)+\sigma+1/2-
{\tilde{R}^2 E \over \tilde{R}^2+\alpha},1+|m|,
\left(1+{\alpha\over \tilde{R}^2}\right)\tilde{r}^2)\right]
e^{im\varphi}\zeta_{\sigma}
\label{24}
\end{equation}
for $E-\sigma-1/2=0,1,2,...$ 
with  $m+\alpha<0$ for $\sigma=+1/2$ and  $m>0$, and 
$m+\alpha<1$ for $\sigma=-1/2$ and $m\le 0$.
The eigenfunctions in (\ref{23}) and (\ref{24})
contain all eigenfunctions of the Hamiltonian by the direct approach. But
(\ref{22}) still allows  further eigenfunctions of Laguerre type
along with their superpartners.
This reduces the vacancy
in the spectra by one $(m+\sigma)$-value
in  comparison to the direct approach.

The other eigenfunctions fulfilling (\ref{22})  are supersinglets.
In order to be normalized as $\tilde{R}\to 0$ they acquire an
$\tilde{R}$-dependent normalization factor that lowers the
leading power in $\tilde{R}$ in (\ref{22}). This very fact ensures
that  fulfilling (\ref{22}) in lowest order is
sufficient for (\ref{22}) to be fulfilled to all orders,
since the higher powers vanish separately.
We call the eigenfunctions with an $\tilde{R}$-dependent
normalization factor
singular eigenfunctions.

For $E=0$ and $\sigma=-1/2$ we find the normalized
singular eigenfunctions for $m+\alpha=1$
$$
\Psi^s_{E=0,\sigma=-1/2,m}(\tilde{r},\varphi)=
{1\over \lambda \sqrt{-\pi\ln \tilde{R}^2}} 
$$
\begin{equation}
\times \left[\theta(\tilde{r}-\tilde{R})
\tilde{r}^{-1}e^{-\tilde{r}^2\over 2}
+\theta(\tilde{R}-\tilde{r}) \tilde{R}^{-1} e^{\alpha /2}
e^{-(1+{\alpha \over \tilde{R}^2}){\tilde{r}^2 \over 2}}
\left({\tilde{r}\over \tilde{R}}\right)^{-m}\right]
e^{im\varphi}\zeta_{-1/2}
\label{25}
\end{equation}
and for $m+\alpha >1$
$$
\Psi^s_{E=0,\sigma=-1/2,m}(\tilde{r},\varphi)=
{\tilde{R}^{m+\alpha-1}\over \lambda
\sqrt{\pi({1 \over m+\alpha-1}+\alpha^{m-1} e^{\alpha}
[\Gamma(-m+1)-\Gamma(-m+1,\alpha)])}}
$$
\begin{equation}
\times \left[\theta(\tilde{r}-\tilde{R})\tilde{r}^{-m-\alpha}
e^{-\tilde{r}^2\over 2}
+\theta(\tilde{R}-\tilde{r}) \tilde{R}^{-m-\alpha} e^{\alpha/2}
e^{-(1+{\alpha \over \tilde{R}^2}){\tilde{r}^2 \over 2}}
\left({\tilde{r}\over \tilde{R}}\right)^{-m}\right]
e^{im\varphi}\zeta_{-1/2}
\label{26}
\end{equation}
with the incomplete gamma function \cite{abr}. We have thus recovered
all the correct eigenfunctions found within the approach in 3.2.

For $E=0$ and $\sigma=+1/2$
we find still another type of singular eigenfunction.
For $m+\alpha=-1$ we obtain
$$
\Psi^s_{E=0,\sigma=+1/2,m}(\tilde{r},\varphi)=
{1\over \lambda \sqrt{-\pi\ln \tilde{R}^2}} 
$$
\begin{equation}
\times \left[\theta(\tilde{r}-\tilde{R})\tilde{r}^{-1}e^{\tilde{r}^2\over 2}
\Gamma(1,\tilde{r}^2)
+\theta(\tilde{R}-\tilde{r}) \tilde{R}^{-1}e^{\tilde{R}^2\over 2}
\Gamma(1,\tilde{R}^2) e^{-\alpha/2}
e^{(1+{\alpha \over \tilde{R}^2}){\tilde{r}^2 \over 2}}
\left({\tilde{r}\over \tilde{R}}\right)^{m}\right]
e^{im\varphi}\zeta_{+1/2}
\label{27}
\end{equation}
and  for  $m+\alpha<-1$
$$
\Psi^s_{E=0,\sigma=+1/2,m}(\tilde{r},\varphi)=
{\tilde{R}^{-m-\alpha-1}\over \lambda
\sqrt{\pi({1 \over -m-\alpha-1}+ e^{-\alpha}(-\alpha)^{-m-1}
[\Gamma(m+1)-\Gamma(m+1,-\alpha)])}} $$
$$
\times \left[\theta(\tilde{r}-\tilde{R})\tilde{r}^{m+\alpha}
e^{\tilde{r}^2\over 2}{\Gamma(-m-\alpha,\tilde{r}^2)\over
\Gamma(-m-\alpha)}\right.
$$
\begin{equation} \left.
+\theta(\tilde{R}-\tilde{r}) \tilde{R}^{m+\alpha} e^{\tilde{R}^2\over 2}
{\Gamma(-m-\alpha,\tilde{R}^2)\over \Gamma(-m-\alpha)}e^{-\alpha/2}
e^{(1+{\alpha \over \tilde{R}^2}){\tilde{r}^2 \over 2}}
\left({\tilde{r}\over \tilde{R}}\right)^{m}\right]e^{im\varphi}
\zeta_{+1/2}.
\label{28}
\end{equation}
They were not anticipated in section 3 by either approach.

The spectrum of $H_{\alpha}$ is displayed in figure 1.

\subsection{The Hilbert space in the regularized coordinate representation}

The scalar product in the regularized coordinate representation is given by
\begin{equation}
\langle \Xi |\Psi \rangle =
\lim_{R\to 0} \int d^2r \, \Xi^*(\vec{r})\Psi(\vec{r}),
\label{29}
\end{equation}
exhibiting the necessary properties.
An important difference between the usual
coordinate representation and the
regularized coordinate representation is that in the
former the null vector is represented by the function that vanishes
everywhere, while in the latter also functions $N(\vec{r})$ that fulfill
\begin{equation}
\lim_{R\to 0} \int d^2r \, N^*(\vec{r})N(\vec{r})= 0 
\label{30}
\end{equation}
also represent the null vector. We restrict the set of functions
representing the null vector to those that fulfill
$\lim_{R \to 0} N(\vec{r})=0$
excluding probability densities that are different from zero on
curves and points in the plane.
Along with the class of functions that represent the null vector, every
state is represented by an equivalence class of functions by addition of
representatives of the null vector.
The corresponding  probability
density defined by $\lim_{R\to 0} \Psi^*(\vec{r}) \Psi(\vec{r}) $
is unique by using  the basic form of a wave function defined below.

The linear space of functions with the above scalar product is
still not a Hilbert space, since we have to ensure completeness; i.~e.\
every Cauchy sequence has to converge.
To this end we specify the wave functions
$\Psi(\vec{r})\in {\cal H}_{\alpha} $.
The numerable set of eigenfunctions of $H_{\alpha}$
(\ref{23})-(\ref{28})
constitutes an orthogonal normalized system with respect
to the scalar product
\begin{equation}
\langle E',\sigma',m'| E ,\sigma, m \rangle =\lim_{R\to 0} \int d^2r \,
\Psi^*_{E',\sigma',m'}(\vec{r})\Psi_{E,\sigma,m}(\vec{r})
=\delta_{E'E}\delta_{\sigma'\sigma}\delta_{m'm}.
\label{31}
\end{equation}
We obtain general wave functions from linear
combinations of the eigenfunctions
\begin{equation}
\Psi(\vec{r})=\sum_{E,\sigma,m}c_{E,\sigma,m}\Psi{}_{E,\sigma,m}(\vec{r})
\quad \mbox{with} \quad
\sum_{E,\sigma,m} c^*_{E,\sigma,m} c_{E,\sigma,m} <\infty.
\label{32}
\end{equation}
This defines the basic form of a wave function as the sum
of eigenfunctions in the basic form.
Wave functions
will always be given in the basic form, unless otherwise indicated.
The space of functions obtained in this way is a
Hilbert space by construction, since we define its elements as limiting
objects of the Cauchy sequence of an increasing number of
members of the sum in (\ref{32}).

The basic form of a wave function $\Xi_{\rm basic}(\vec{r})$
is obtained from  an arbitrary equivalent wave function
$\Xi_{\rm non-basic}(\vec{r})$ by
\begin{equation}
\, \Xi_{\rm basic}(\vec{r})=
\sum_{E,\sigma,m}\Psi_{E,\sigma,m}(\vec{r}) \lim_{R \to 0 } \int  d^2r'\,
\Psi^*_{E,\sigma,m}(\vec{r} \, {}')\Xi_{\rm non-basic}(\vec{r}\, {}').
\label{34}
\end{equation}
This relation also expresses the fact that the eigenfunctions of the
Hamiltonian supply a resolution of unity. It is this
requirement that enforces the construction
of the Hilbert space as presented here.

For $\alpha=0$ the Hilbert space is clearly $L_2 \otimes \bbbc^2$. 
Also for non-integer $\alpha\ne 0$
the eigenfunctions of Laguerre type provide a resolution of unity for
$L_2\otimes \bbbc^2$ \cite{sze}.
For $|\alpha|\ge 1$ the singular eigenfunctions are added.
For integer $\alpha \ne 0$ from the functions of Laguerre type those
with $(m+\sigma)$ of the vacancy are removed.

The projection of wave functions
$c_{E,\sigma,m}=\langle E,\sigma,m|\Psi  \rangle $
in the regularized coordinate representation on the elements of the
complete orthogonal system yields a matrix representation where each
state is uniquely represented by the vector
$(\{c_{E,\sigma,m} \})$, e.~g.\ the null vector is given by $(0,0,...)$.

The pairing of states by supersymmetry is possible
in the regularized coordinate representation
only  by the freedom of adding a  representative of the null vector.
E.~g.\ the supersymmetric pairing
of  $\Psi_{E,\sigma=-1/2,m}(\vec{r})$
with $\Psi_{E,\sigma=+1/2,m-1}(\vec{r})$ may be expressed by 
$$
\lim_{R \to 0}
\int d^2r \, \Xi^*(\vec{r})
Q_{\alpha} \Psi_{E,\sigma=-1/2,m}(\vec{r})=
\lim_{R \to 0}
\int d^2r \, \Xi^*(\vec{r})
[Q_{\alpha} \Psi_{E,\sigma=-1/2,m}(\vec{r})+N(\vec{r})]
$$
\begin{equation}
= \lim_{R \to 0}
\int d^2r \, \Xi^*(\vec{r})
(\pm \sqrt{E}) \Psi_{E,\sigma=+1/2,m-1}(\vec{r}),
\label{35}
\end{equation}
with the upper sign for eigenfunctions from (\ref{23}) and
the lower sign  for eigenfunctions from (\ref{24}).
The function $Q_{\alpha} \Psi_{E,\sigma=-1/2,m}(\vec{r})$
with a step  at $r=R$ is repaired in order to  
obtain $\Psi_{E,\sigma=+1/2,m-1}(\vec{r})$ in the basic form by 
addition of the appropriate $N(\vec{r})$.

\subsection{Probability densities}

At this point it is useful to discuss the modulus squared
$\Psi^*(\vec{r}) \Psi(\vec{r}) $ in the limit $R\to 0$
related to measurement as a probability density for 
the eigenfunctions of the Hamiltonian.

Of course, for the eigenfunctions of Laguerre type the functional
form of the exterior part of the modulus squared 
extends to all $r\ne 0$ if $R \to 0$.
So only the single point $r=0$ remains to be considered.

For $|\alpha| \ge 1$ we find that the probability density is
zero at $r=0$.
The vast majority of the
eigenfunctions of Laguerre type vanish as $r \to 0$,
yielding  a  density that is continuous everywhere.
We have a non-continuous behavior at $r=0$
if the outside density reaches a constant value
or diverges as $r\to 0$, since the probability density
is still zero at $r=0$. Thus, for $|\alpha| \ge 1$ the probability densities
of the non-singular wavefunctions built on
eigenfunctions of the Laguerre type avoid the location of the flux tube.

By contrast, for $|\alpha|<1$ the value at $r=0$ agrees with
the behavior of the exterior part as $r\to 0$,
such that either the probability density
is continuous everywhere or the value at $r=0$
diverges as $R \to 0$ along with
the exterior part  diverging  as $r\to 0$.

The normalization of the singular eigenfunctions implies
\begin{equation}
\lim_{R\to 0}\int_{\cal G}d^2r \Psi^{s*}_{E,\sigma,m}(\vec{r})
\Psi^{s}_{E,\sigma,m}(\vec{r})=1
\label{35a}
\end{equation}
for any ${\cal G}$ containing $r=0$. If  ${\cal G}$ did not contain
$r=0$, the integrand would converge uniformly to zero and
the integral would vanish as $R\to 0$.
This means that the modulus squared of a singular eigenfunction for
finite $R$ is a delta convergent function \cite{gel}, i.~e.\
\begin{equation}
\lim_{R\to 0}
\Psi_{E=0,\sigma,m}^{s*}(\vec{r}) \Psi^s_{E=0,\sigma,m}(\vec{r})
=\delta^2(\vec{r}),
\label{36}
\end{equation}
where the limit $R \to 0$ is to be taken outside of an integral.
Thus, the probability density  of  a singular eigenfunction is that
of a two dimensional point particle, concentrated at  the location of the
flux tube.

Singular eigenfunctions occur for $|\alpha| \ge 1$ just when 
the probability densities of non-singular wavefunctions avoid the flux tube
as mentioned above.
Although not being formally the reason,
the fact that the probability density of the singular eigenfunctions
and the eigenfunctions of Laguerre type are attributed to disjoint regions
in the plane conforms nicely with the mutual orthogonality.
Indeed the singular eigenfunctions are orthogonal to any function
$\Xi(\vec{r})$ yielding a
Taylor expansion about $r=0$
\begin{equation}
\lim_{R\to 0} \int d^2r \, \Psi_{E=0,\sigma,m}^{s*}(\vec{r}) \Xi(\vec{r})=0.
\label{36a}
\end{equation}
From this we also conclude that 
$\Psi^s_{E=0,\sigma,m}(\vec{r})$ itself neither is a delta convergent
function nor does it converge to a derivative of the
delta function. Since this exploits all possibilities
for generalized functions concentrated at a point \cite{gel}, the
corresponding functional must be zero.

We define a general singular wave function
$\Psi^s(\vec{r})$ as a linear combination
of singular eigenfunctions. It also has the properties
(\ref{36}) and (\ref{36a}) if it is normalized.

\subsection{Classification  of the Hilbert spaces ${\cal H}_{\alpha}$}

By $\alpha $ a sequence of Hilbert spaces is parametrized. We have
determined 
each Hilbert space by solving the
eigenvalue problem of the
Hamiltonian exactly for all  values of $\alpha $. It is,
therefore, not necessary to resort to
perturbation theory.
Still, in view of extensions of the system and analogous situations the
question arises whether
we can get arbitrarily close to the exact solution of a system with
$\alpha '=\alpha+\Delta \alpha$ by
perturbation theory in $\Delta \alpha$
if the system with $\alpha $ is exactly known.

A necessary condition for this is that exact  solutions of the
perturbed system can be expanded in terms of the exact solutions
of the unperturbed system. An exact solution of the
perturbed system that is orthogonal to all exact solutions of the
unperturbed system can never be reached by a perturbation series.

The real numbers that $\alpha $ can assume decompose into
the isolated points  $\pm 1;\pm 2;...$
and the complementary open intervals $ (\! -\! 1,1);$
$(1,2)$, $ (2,3),...$; $(\! - \! 2,\! - \! 1)$, $ (\! -\! 3,\! -\! 2),...$ 
This is  a result of the occurrence of the singular states and
the vacancy in the $(m+\sigma)$-range for $\alpha =\pm 1, \pm 2, ...$
These sets represent equivalence classes with respect to
the relation ${\cal H}_{\alpha} ={\cal H}_{\alpha'}$.
For any $\alpha $ and $\alpha '$
in different equivalence classes we have
${\cal H}_{\alpha} \ne {\cal H}_{\alpha'}$
such that there is at least one state
in ${\cal H}_{\alpha}$ or ${\cal H}_{\alpha'}$
that is orthogonal to all states of the other.
For any two neighboring equivalence classes it turns out that
both Hilbert spaces contain at least one state
that is orthogonal to all states of the other.
For these states perturbation theory fails
and the above equivalence classes constitute perturbatively disjoint
sectors. We summarize:
\begin{trivlist}

\item[] i) As expected,
$\Delta \alpha$ may not be arbitrarily large for a successful application of
perturbation theory; in fact $|\Delta \alpha|<2$.

\item[] ii) However, very small $\Delta \alpha$ does not guarantee
successful application of perturbation theory, since $\alpha $ and
$\alpha +\Delta \alpha$ could still be in different equivalence classes.

\item[] iii) The open interval $(\! -\! 1,1)$ is distinguished as
the interval length is exceptionally $2$, while for the other
equivalence classes we have interval length of $1$ or $0$.
$\alpha=0$ is the only integer that is useful as a perturbative
starting point for its neighboring values.
The Hilbert space is that of square-integrable 2-spinors
${\cal H}_{\alpha}=L_2\otimes \bbbc^2$ \cite{sze}.
\end{trivlist}

The dependence
of $H_{\alpha}$ and the superoperators on $\alpha $ does not anticipate the
decomposition into disjoint perturbative sectors. However,
by $\alpha $ we parametrize magnetic flux, which is the global topological
invariant associated with a $U(1)$ principal fiber bundle
characterizing the topology of the bundle \cite{egu}.
Thus, the discontinuous behavior of the Hilbert spaces
reflects the differing topologies, i.~e.\
discontinuous geometries, as $\alpha$ is varied.

\subsection{An index for the singular flux tube }

Index theorems relate the spectrum of a differential operator on a  manifold
to a global topological invariant. The number of
eigenstates of $E=0$ with
$\sigma=-1/2$ minus those with  $\sigma=+1/2$
yield an index,
referred to as the Witten index
\cite{wit}\cite{wit2} in supersymmetric systems. It corresponds
to the index of the euclidian Dirac operator in two dimensions,
which is subject to index theorems.
The Atiyah-Singer index theorem \cite{as} for manifolds without boundary
and the Atiyah-Patodi-Singer \cite{aps} index theorem for manifolds with
boundary apply to compact manifolds.
For the $U(1)$ case in two dimensions 
the Atiyah-Singer index theorem reduces to the statement
that the above index equals the total flux in units of
the elementary flux quantum.
The index theorems are not applicable to our system
as the plane is not a compact manifold.
Both the above index and the magnetic flux through the plane are
infinite due to the presence of the homogeneous magnetic field.

More can be said in view of the fact that in the limit $R\to 0$
both the eigenfunctions of $E=0$  and the flux may be attributed
to disjoint subspaces of the plane: the point $r=0$ and its complement.

For $r\ne 0$ the eigenfunctions of Laguerre type with $E=0$ and
$\sigma=-1/2$
that are not localized close to the flux tube can be attributed to
flux quanta of the homogeneous field as for $\alpha =0$ \cite{lan}.

For $r=0$ we define the index $I^s(\alpha)$
as the number of singular eigenfunctions
with $\sigma=-1/2$ minus the number of singular
eigenfunctions with $\sigma=+1/2$.
If $[\alpha]$ denotes the next integer below $\alpha$,
it is given by
\begin{equation}
I^s(\alpha)=[\alpha]\theta(\alpha)-[-\alpha]\theta(-\alpha),
\label{62}
\end{equation}
displayed in  figure 2.
For integer $\alpha $ we have $I^s(\alpha)=\alpha$,
as if we had a compact manifold as in \cite{as}.

\section{Explicit breaking of supersymmetry by $g\ne 2$}

Concerning the applicability of the results  obtained so far, we are
facing the fact that already in vacuo $g \approx 2.0023$ by
QED corrections. Even by this tiny deviation from $g=2$ supersymmetry is
explicitly broken. The question is to what extent our results remain valid
for $g\ne 2$. We discuss only the effect of QED corrections here,
but for other perturbative background effects
the arguments might be analogous.

The deviation of $g$ from $2$ is calculated under the assumption of
a homogeneous field that is sufficiently weak, i.~e.\
$|-\vec{\mu} \cdot \vec{B} |\ll Mc^2$ \cite{new1}\cite{jan}\cite{sch}.
For the outside region we can assume this to be true.
By contrast, inside of the
flux tube the magnetic field grows as $R\to 0$. Assuming that
$R\to 0$ idealizes $R\approx |\alpha|^{1/2}\lambda_C$, we have
$|-\vec{\mu} \cdot \vec{B}|\approx Mc^2$ and it is not justified to put
$g\ne 2$ there. The most reasonable choice
is to maintain $g=2$ within the flux tube.
This may also be rephrased heuristically
following the correspondence principle.
A strong magnetic field implies high occupation numbers
of the quantum electromagnetic field. The limit of high occupation
numbers corresponds to classical behavior.
As a result, contributions to QED processes from the fluctuations
of the  quantum electromagnetic field are suppressed.

The most obvious implementation of $g=2$ inside and $g\ne 2$ outside
would be to couple the entire magnetic field with $g=2$ for $r\le R$ and
with $g\ne 2$ for $r>R$. The result would, however, be inconsistent with our
regularization. On the one hand, for $|\alpha|>1$ the expectation values
of the Hamiltonian
with the singular eigenfunctions would  not reproduce the
eigenvalue. On the other hand, the regularized Hamiltonian
would not yield the Hamiltonian of the
homogeneous field for $\alpha= 0$.
Therefore, we proceed differently.

We couple $B\vec{e}_z$ by $g=2(1+\kappa) \ne 2$ everywhere and
$(\alpha\Phi/\pi R^2)\vec{e}_z$ by $g=2$ inside of the flux tube, 
which amounts to a term
$(1+\kappa)B S_z + \theta(R-r)(\alpha\Phi/\pi R^2) S_z$
in the Hamiltonian.
$\alpha\Phi / \pi R^2$ dominates $B$
in the limit $R\to 0$ and we effectively
attribute to the interior $g=2$ and to the exterior  $g\ne 2$ in a
consistent way. The new Hamiltonian is
\begin{equation}
H_{\alpha}^{\kappa}=H_{\alpha}+\kappa S_z=
\{Q_{\alpha},Q_{\alpha}^{\dagger}\}+\kappa S_z,
\label{63}
\end{equation}
entailing the explicit breaking of supersymmetry
\begin{equation}
[H_{\alpha}^{\kappa},Q_{\alpha}]=\kappa Q_{\alpha} \qquad \mbox{and} \qquad
[H_{\alpha}^{\kappa},Q_{\alpha}^{\dagger}]=-\kappa Q_{\alpha}^{\dagger}.
\label{64}
\end{equation}
A lower bound for the expectation values of $H^{\kappa}_{\alpha}$ is now
$\langle \Psi |H_{\alpha}^{\kappa}|\Psi \rangle \ge -\kappa/2 $
instead of zero.
For the solution of the eigenvalue problem of $H^{\kappa}_{\alpha}$
we have still $[H^{\kappa}_{\alpha},L_z]=[H^{\kappa}_{\alpha},S_z]=0$.
The matching condition (\ref{22}) is not affected, since the
leading contributions at small $R$ of both the
outside and the inside parts of (\ref{14}) are independent of $\kappa$.
We obtain the eigenfunctions for $\kappa\ne 0$ from those
for $\kappa=0$
given by (\ref{23})-(\ref{28}) replacing $E$ by $E^{\kappa}-\kappa \sigma$.
Thus as functions of $\vec{r}$ they are unchanged, while  
the eigenvalues
$E^{\kappa}=E+\kappa\sigma$
of $H^{\kappa}_{\alpha}$ are shifted.

The eigenfunctions with $E^{\kappa}=\kappa\sigma$ are now the supersinglets.
The other eigenfunctions are still paired by $Q_{\alpha}$ and
$Q_{\alpha}^{\dagger}$, although having
different energy eigenvalues.
One might object that, since the
superoperators no longer commute with the Hamiltonian,
the pairing is not enforced as for the $\kappa =0$ case, where the pairing
has effectively determined the correct ``boundary conditions''.
Then additional unpaired eigenfunctions
could enter the spectrum for $\kappa\ne 0$.
This is not true, however. Additional eigenfunctions would have to disappear
abruptly in
the limit $\kappa \to 0$. Since $\kappa\ne 0$ is a perturbative
effect, one has to impose a continuous limit $\kappa \to 0$.

\section{The singular Aharonov-Bohm system}

The special value of $B=0$ has not been covered so far,
as for $r\to \infty$
we can no longer impose the vanishing of the eigenfunctions
of the Hamiltonian. By the discussion of the last section
only the case of $g=2$ needs to be considered here.

Lacking a knowledge of the behavior of the eigenfunctions as $r \to \infty$,
we have to start with a more fundamental criterion for selecting the correct
eigenfunctions. Ultimately, the set of
correct eigenfunctions of the Hamiltonian
has to supply a resolution of unity in the Hilbert space.

We consider the special case $\alpha =0$,
where this system reduces to the free particle in two dimensions.
The corresponding eigenfunctions  are plane waves, not actually constituting
a normalized orthogonal set.
Still these eigenfunctions supply a resolution of
unity in the Hilbert space of
square-integrable 2-spinors ${\cal H}_0=L_2 \otimes \bbbc^2$
by theorems on the Fourier transform.
Obviously, the behavior of the system for $r\to \infty$
does not affect the Hilbert space.
A normalizable orthogonal system spanning ${\cal H}_0=L_2 \otimes \bbbc^2$
is given by the eigenfunctions of the
Hamiltonian of the homogeneous magnetic field.

This can be generalized to $\alpha \ne 0$.
We demand that the solutions of the
eigenvalue problem of $H_{\alpha}^{AB}$ at $B=0$ yield a resolution of unity
in the Hilbert space ${\cal H}_{\alpha}$ spanned 
by the normalized orthogonal set of eigenfunctions
(\ref{23})-(\ref{28}) at $B\ne 0$.
For $\alpha \ne 0$ the system looks the same as for  $\alpha=0$
as $r \to \infty$.
Generally, the value of $B$ by
which the systems might differ as $r \to \infty$
does not affect the Hilbert space.
But we have shown 
that the  Hilbert space underlying
the system does depend on the value of $\alpha $,
i.~e.\ on the behavior at the origin.

According to 4.4 any system with $B\ne 0$ and
$\alpha'$ within the same equivalence class as
$\alpha$ would supply a suitable reference basis.
We choose $\alpha = \alpha'$ for convenience.


The operators
$$
H_{\alpha}^{AB}=-{1\over 4}\left({1 \over r}{\partial \over \partial r}
r {\partial \over \partial r}+ {1 \over r^2} {\partial^2 \over \partial
\varphi^2}\right)
$$
\begin{equation}
+\theta(r-R)\left(
-{i\alpha \over 2 r^2}{\partial \over \partial \varphi}+
{\alpha^2 \over 4r^2}\right)
+\theta(R-r){\alpha \over R^2}\left({\alpha\over 4R^2}r^2
+ {1 \over 2} r^2 -
{i \over 2}{\partial \over \partial \varphi}+S_z \right),
\label{65}
\end{equation}
$$
Q_{\alpha}^{AB}=S_+ V_{\alpha}^{AB}=
S_+{e^{-i\varphi} \over 2}\left[{\partial \over \partial r }
-{i\over r}
{\partial \over \partial \varphi } +{\alpha \over r}
+\alpha  \theta(R-r)\left({r \over R^2} -{1 \over r}\right)\right],
$$
\begin{equation}
(Q_{\alpha}^{AB})^{\dagger}=S_-(V_{\alpha}^{AB})^{\dagger}=
S_-{e^{i\varphi} \over 2}\left[-{\partial \over \partial r }
-{i\over r}{\partial \over \partial \varphi } +{\alpha \over r}
+\alpha\theta(R-r)\left({r \over R^2} -{1 \over r}\right)\right]
\label{66}
\end{equation}
are obtained by multiplication of
(\ref{12}) and (\ref{13}) with
$\lambda^{-2}$ and $ \lambda^{-1}$, respectively,
and the subsequent limit $\lambda \to \infty$.
These operators are now measured in units of
$2\hbar^2 /M$ and $2^{1/2}\hbar /(M)^{1/2}$, respectively.
The supersymmetry  algebra (\ref{2}) is fulfilled by
$Q_{\alpha}^{AB}$,
$(Q_{\alpha}^{AB})^{\dagger}$
and $H_{\alpha}^{AB}$.

We solve the eigenvalue problem of  the Hamiltonian in the form
\begin{equation}
\langle E,\sigma,m|\left(H_{\alpha}^{AB}-{k^2\over 4}\right)|k\rangle =0.
\label{67}
\end{equation}
A necessary condition on the solutions to yield a resolution of unity
is that they have to be within the
Hilbert space; i.~e., all $\langle E,\sigma,m|k\rangle$
must be finite  and at least one for a given $|k\rangle$ must be different
from zero.
Hermiticity of the Hamiltonian
imposes a further condition on the solutions, which requires
orthogonality $\langle k'|k \rangle=0$ if $k\ne k'$.
Due to $[H_{\alpha}^{AB},L_z]=[H_{\alpha}^{AB},S_z]=0$, we use the ansatz
$\Psi_{k,\sigma,m}(r,\varphi)= \psi_{k,\sigma,m}(r)
e^{im\varphi}\zeta_{\sigma}$.
Inside of the flux tube we have again to solve a Kummer equation.
Outside of the flux tube we have a Bessel equation for $k>0$, which
reduces to the Laplace equation for $k=0$.
If we impose continuity at $r=R$, from the eigenvalue equation (\ref{67})
the matching condition (\ref{22}) follows again.
We will not  combine the two linear independent solutions
for $k>0$ to yield differentiable solutions at $r=R$, such that
(\ref{22}) would be fulfilled even before taking the limit $R \to 0$.
Instead, in order to proceed coherently with
the case of the additional homogeneous field we
only demand continuity.
The discussion based on the leading behavior of the solutions in section 4.1
is repeated literally by using only one of the
linear independent solutions for $r>R$.
The result is for $k>0$
$$
\Psi_{k,\sigma,m}(r,\varphi)= 
\left[\theta(r-R) J_{m+\alpha}(kr)
\vphantom{\tilde{R}^2 E \over \tilde{R}^2} \right.
+\theta(R-r) {R^{-|m|}e^{\alpha\over 2} J_{m+\alpha}(kR)\over
{}_1F_1(m\theta(m)+\sigma+1/2-{R^2 k^2 \over 4\alpha},1+|m|,\alpha)}
$$
\begin{equation}\left.
\times e^{-\alpha r^2/2R^2} r^{|m|}
{}_1F_1(m\theta(m)+\sigma+1/2-{R^2 k^2 \over 4\alpha},1+|m|,
{\alpha\over R^2}r^2)\right]e^{im\varphi}\zeta_{\sigma}
\label{68}
\end{equation}
with  $m+\alpha>0$ for $\sigma=-1/2$ and  $m<0$, and 
$m+\alpha>-1$ for $\sigma=+1/2$ and $m\ge 0$;
$$
\Psi_{k,\sigma,m}(r,\varphi)=
\left[\theta(r-R) J_{-m-\alpha}(kr)
\vphantom{\tilde{R}^2 E \over \tilde{R}^2}\right.
+\theta(R-r) {R^{-|m|} e^{\alpha\over 2}J_{-m-\alpha}(kR) \over
{}_1F_1(m\theta(m)+\sigma+1/2-{R^2 k^2\over 4\alpha},1+|m|,\alpha)}
$$
\begin{equation}\left.
\times e^{-\alpha r^2/2R^2} r^{|m|}
{}_1F_1(m\theta(m)+\sigma+1/2-{R^2 k^2 \over 4\alpha},
1+|m|,{\alpha\over R^2}r^2)\right] e^{im\varphi}\zeta_{\sigma}
\label{69}
\end{equation}
with $m+\alpha<0$ for $\sigma=+1/2$ and $m>0$, and 
$m+\alpha<1$ for $\sigma=-1/2$ and $m\le 0$.

For $k=0$ the solutions are
$$\Psi^s_{k=0,\sigma=\mp1/2,m}(r,\varphi)
$$
\begin{equation}
=\nu[\theta(r-R){r }^{\mp(m+\alpha)} +\theta(R-r) {R }^{\mp \alpha}
e^{\pm \alpha/2} e^{\mp \alpha r^2/2R^2} {r}^{\mp m}]
e^{im\varphi}\zeta_{\mp1/2}
\label{70}
\end{equation}
for $\sigma=\mp 1/2$, $\pm m\le 0$ and $\mp (m+\alpha) >1/2$.
There are three types of solutions for $k=0$ concerning their
normalization behavior.
\begin{trivlist}

\item[] i) We have
$\nu= {R }^{\pm(m+\alpha)-1}\pi^{-1/2}[(\pm (m+\alpha)-1)^{-1}
+(\pm \alpha)^{\pm m-1} e^{\pm \alpha}\gamma(\mp m+1,\pm \alpha)]^{-1/2}$
for $\mp (m+\alpha)>1$ and $\sigma=\mp 1/2$.
The normalized eigenfunctions
can be found by directly performing the limit of the
vanishing homogeneous field, i.~e.\ $\lambda \to \infty$ in the normalized
eigenfunctions in (\ref{26}) and (\ref{28}).
The modulus squared is still a delta convergent function.

\item[] ii) We have
$\nu=[-\pi\ln R^2]^{-1/2}$
for $\mp (m+\alpha)=1$ and $\sigma=\mp 1/2$.
The eigenfunctions occur for integer $\alpha \ne 0$ and they correspond
to the only eigenfunction in the $(m+\sigma)$-vacancy in the case of the
additional homogeneous magnetic field. However, they are not obtained
by the limit $\lambda \to \infty $ from the normalized functions (\ref{25})
and (\ref{27}), since they yield a logarithmically divergent norm.
$\nu$ is determined by demanding that
$\langle \Psi^{s,\lambda}_{E=0,\sigma,m}
|\Psi^{s,\lambda'=\infty}_{k=0,\sigma,m}\rangle=1 $
for arbitrary $\lambda\ne \infty$.
Still as for finite $\lambda$, the modulus squared yields a delta convergent
function, since the weak logarithmic divergent behavior as $r\to \infty $
is compensated by any decaying test function.

\item[] iii) For $1>\mp (m+\alpha)>1/2$ and $\sigma=\mp 1/2$ we
put $\nu=1$ for convenience.
These solutions share their property of not being normalizable with the
continuum states with $k>0$.
\end{trivlist}

The eigenfunctions (\ref{68})-(\ref{70}) constitute a resolution
of unity in ${\cal H}_{\alpha}$.
This follows, on the one hand,
from the fact that the eigenfunctions of the free
particle represent a resolution of unity in $L_2 \otimes \bbbc^2$.
In the case of integer $\alpha \ne 0$ both for finite
$\lambda$ and $\lambda \to
\infty $ the vacancy occurs for the same $(m+\sigma)$-value.
On the other hand, the singular eigenfunctions for $\lambda \to \infty$
occur for the
same quantum numbers $m$ and $\sigma$ as for finite $\lambda$.
They yield $\langle \Psi^{s,\lambda'=\infty}_{\sigma',m'}
|\Psi^{s,\lambda}_{\sigma,m} \rangle
=\delta_{m,m'}\delta_{\sigma,\sigma'}$ for all finite $\lambda$ and,
therefore, in the finite subspace of the singular wave  functions
we have a resolution of unity.
Since the singular eigenfunctions and the non-singular ones
are mutually orthogonal
by (\ref{36a}), the corresponding  resolutions of unity
do not interfere and the
sum of both yields a resolution of unity of the entire
${\cal H}_{\alpha}$.

Having only a finite number of eigenfunctions with $k=0$, we define
$I^{AB}(\alpha)$ as the number of eigenfunctions
for $k=0$ and $\sigma=-1/2$
minus those for $k=0$ and $\sigma=+1/2$. It is given by
\begin{equation}
I^{AB}(\alpha)=
\left\{
\begin{array}{lll}
n & \mbox{for} \,\, \alpha \in   \, [n-1/2,n+1/2) &
\quad n=-1,-2,...\\
0 & \mbox{for} \,\, \alpha \in [-1/2,1/2]     & \\
n & \mbox{for} \,\, \alpha \in  (n-1/2,n+1/2] &
\quad n=1,2,...
\end{array} \right.
\label{79}
\end{equation}
sketched in figure 2 on the right hand side.
$I^s(\alpha)$ is the same as for the
nonvanishing homogeneous field. 
(\ref{79}) coincides with the Atiyah-Patodi-Singer index theorem for
compact manifolds with boundary, according to \cite{for} by use of the
appropriate boundary conditions. For integer $\alpha $ both
$I^s(\alpha)$ and $I^{AB}(\alpha)$ give the same value $\alpha$ in agreement
with the Atiyah-Singer index theorem \cite{as} valid on compact
manifolds without
boundaries, which is plausible since in this case a compactification is
possible.

\section{Concluding remarks}

Our results do not depend on our special arrangement of the magnetic field,
i.~e.\ regularization. Smearing out the discontinuity of the magnetic field
between $R-\eta$ and $R$ adds
to the matching condition (\ref{22}) a correction of $O(\eta)$
that is forced to vanish
by the limit $R\to 0$.

The need for a regularization is dictated by consistency. And for this
reason also the consequences of the use of a regularization
are to be taken seriously.
Ultimately, in what way these idealized structures could be 
realized in nature is yet another question to be resolved by experiment.

In view of the quantum Hall effect
for $\alpha =1,2,...$ the eigenfunctions
of Laguerre type preserve the structure of
the lowest Landau level of the homogeneous field $\alpha =0$ supplying
the states to be occupied for the
incompressible quantum fluid.
The singular states are orthogonal to these, thus not participating.
If the singular states remained unoccupied, the
number of electrons per non-singular flux quantum
would be the same as for $\alpha=0$.
However, the total magnetic flux would be
increased by the contribution of the
singular flux tube.
In other words if there were a singular flux tube without occupation of the
corresponding singular states, the filling fraction
would be effectively decreased.

Finally, this system might be interesting
as a toy model for supersymmetric field theories, for two reasons. 
First, because the singular
flux parametrizes different topological situations and thereby controls
the subspace of supersinglets, related to 
spontaneous breaking of supersymmetry.
Second,  the deviation from $g=2$ provides
an explicit breaking of supersymmetry from effective physics
such as QED corrections
at lower scales
$\lambda_C$ than the scale $\lambda$ of the quantum mechanical system.

\vskip 1.0truecm
\noindent {\bf Acknowledgment:}

\noindent I am indebted to D.\ Schiller and H.\ D.\ Dahmen
for helpful discussions.

\vfill

\newpage

\setlength{\unitlength}{.35cm}


\vfill 

{\small \noindent For caption see next page. }
\newpage

{\small
\noindent FIG. 1 (previous page). Spectrum of  $H_{\alpha}$

\noindent
A state of Laguerre type with $\sigma=-1/2$
is indicated by 
\setlength{\unitlength}{.4cm}
%
and a state of Laguerre type with $\sigma=+1/2$ by
\setlength{\unitlength}{.4cm}
%
.
A supersymmetric pair of such states
is indicated by 
\setlength{\unitlength}{.4cm}
%
.
Singular states with $\sigma=-1/2$ are indicated by
\setlength{\unitlength}{.4cm}
%
and  singular states with  $\sigma=+1/2$ by
\setlength{\unitlength}{.4cm}
%
.
For $\alpha =0$ the eigenvalues of $H_{\alpha}$ constitute
a regular rectangular lattice within a $3/8$-sector of the
$E$-$(m+\sigma)$-plane.
The rectangular lattice acquires a line defect for $\alpha \ne 0$.
For integer $\alpha \ne 0$ the defect
is a vacancy line between the left and
the right part of the spectrum, i.~e.\
one $(m+\sigma)$-value  is missing, entailing the absence
of the corresponding Fourier components $\sim e^{im\varphi}$
among the eigenfunctions of the Hamiltonian.
For non-integer $\alpha $ there is a vertical shift
of the right block  against the left block by the non-integer
part of $\alpha$.
All non-integer values of 
$\alpha $ between two integers show the same
qualitative pattern differing only in the amount of 
vertical shift of the right block.
The spectrum is given for half-integer values.
}

\newpage

\setlength{\unitlength}{.8cm}
\begin{picture}(17,8)(-3,-4)

\begin{picture}(8,8)(0,0)

\put (0,-2.92) {\line(0,1){.17}}\put (0,-2.58) {\line(0,1){.17}}
\put (0,-2.25) {\line(0,1){.17}}

\put (0,-1.92) {\line(0,1){.17}}\put (0,-1.58) {\line(0,1){.17}}
\put (0,-1.25) {\line(0,1){.17}}

\put (0,-0.92) {\line(0,1){.17}}\put (0,-0.58) {\line(0,1){.17}}
\put (0,-0.25) {\line(0,1){.17}}

\put (0,0.08) {\line(0,1){.17}}\put (0,0.42) {\line(0,1){.17}}
\put (0,0.75) {\line(0,1){.17}}

\put (0,1.08) {\line(0,1){.17}}\put (0,1.42) {\line(0,1){.17}}
\put (0,1.75) {\line(0,1){.17}}

\put (0,2.08) {\line(0,1){.17}}\put (0,2.42) {\line(0,1){.17}}
\put (0,2.75) {\line(0,1){.17}}

\put (0,3.08) {\vector(0,1){.17}}

\put (-2,3) {$I^{s}(\alpha)$}

\put (-3.92,0) {\line(1,0){.17}}\put (-3.58,0) {\line(1,0){.17}}
\put (-3.25,0) {\line(1,0){.17}}

\put (-2.92,0) {\line(1,0){.17}}\put (-2.58,0) {\line(1,0){.17}}
\put (-2.25,0) {\line(1,0){.17}}

\put (-1.92,0) {\line(1,0){.17}}\put (-1.58,0) {\line(1,0){.17}}
\put (-1.25,0) {\line(1,0){.17}}

\put (-0.92,0) {\line(1,0){.17}}\put (-0.58,0) {\line(1,0){.17}}
\put (-0.25,0) {\line(1,0){.17}}

\put (0.08,0) {\line(1,0){.17}}\put (0.42,0) {\line(1,0){.17}}
\put (0.75,0) {\line(1,0){.17}}

\put (1.08,0) {\line(1,0){.17}}\put (1.42,0) {\line(1,0){.17}}
\put (1.75,0) {\line(1,0){.17}}

\put (2.08,0) {\line(1,0){.17}}\put (2.42,0) {\line(1,0){.17}}
\put (2.75,0) {\line(1,0){.17}}

\put (3.08,0) {\line(1,0){.17}}\put (3.42,0) {\line(1,0){.17}}
\put (3.75,0) {\line(1,0){.17}}

\put (4.08,0) {\vector(1,0){.17}}

\put (4,-1) {$\alpha $}

\put (1,-.1) {\line(0,1){.2}}\put (1,-1) {$1$}

\put (-.1,1) {\line(1,0){.2}}\put (-1,1) {$1$}

\put (1,1){\circle*{.1}}\put (2,2){\circle*{.1}}
\put (-1,-1){\circle*{.1}}\put (-2,-2){\circle*{.1}}

\thicklines

\put (-3,-2) {\line(1,0){1}}\put (-2,-1) {\line(1,0){1}}
\put (-1,0) {\line(1,0){2}}\put (1,1) {\line(1,0){1}}
\put (2,2) {\line(1,0){1}}

\end{picture}

\begin{picture}(8,8)(-1,0)

\put (0,-2.92) {\line(0,1){.17}}\put (0,-2.58) {\line(0,1){.17}}
\put (0,-2.25) {\line(0,1){.17}}

\put (0,-1.92) {\line(0,1){.17}}\put (0,-1.58) {\line(0,1){.17}}
\put (0,-1.25) {\line(0,1){.17}}

\put (0,-0.92) {\line(0,1){.17}}\put (0,-0.58) {\line(0,1){.17}}
\put (0,-0.25) {\line(0,1){.17}}

\put (0,0.08) {\line(0,1){.17}}\put (0,0.42) {\line(0,1){.17}}
\put (0,0.75) {\line(0,1){.17}}

\put (0,1.08) {\line(0,1){.17}}\put (0,1.42) {\line(0,1){.17}}
\put (0,1.75) {\line(0,1){.17}}

\put (0,2.08) {\line(0,1){.17}}\put (0,2.42) {\line(0,1){.17}}
\put (0,2.75) {\line(0,1){.17}}

\put (0,3.08) {\vector(0,1){.17}}

\put (-2,3) {$I^{AB}(\alpha)$}

\put (-3.92,0) {\line(1,0){.17}}\put (-3.58,0) {\line(1,0){.17}}
\put (-3.25,0) {\line(1,0){.17}}

\put (-2.92,0) {\line(1,0){.17}}\put (-2.58,0) {\line(1,0){.17}}
\put (-2.25,0) {\line(1,0){.17}}

\put (-1.92,0) {\line(1,0){.17}}\put (-1.58,0) {\line(1,0){.17}}
\put (-1.25,0) {\line(1,0){.17}}

\put (-0.92,0) {\line(1,0){.17}}\put (-0.58,0) {\line(1,0){.17}}
\put (-0.25,0) {\line(1,0){.17}}

\put (0.08,0) {\line(1,0){.17}}\put (0.42,0) {\line(1,0){.17}}
\put (0.75,0) {\line(1,0){.17}}

\put (1.08,0) {\line(1,0){.17}}\put (1.42,0) {\line(1,0){.17}}
\put (1.75,0) {\line(1,0){.17}}

\put (2.08,0) {\line(1,0){.17}}\put (2.42,0) {\line(1,0){.17}}
\put (2.75,0) {\line(1,0){.17}}

\put (3.08,0) {\line(1,0){.17}}\put (3.42,0) {\line(1,0){.17}}
\put (3.75,0) {\line(1,0){.17}}

\put (4.08,0) {\vector(1,0){.17}}

\put (4,-1) {$\alpha $}

\put (1,-.1) {\line(0,1){.2}}
\put (1,-1) {$1$}

\put (-.1,1) {\line(1,0){.2}}
\put (-1,1) {$1$}

\put (.5,0){\circle*{.1}}\put (1.5,1){\circle*{.1}}
\put (-.5,0){\circle*{.1}}
\put (-1.5,-1){\circle*{.1}}\put (-2.5,-2){\circle*{.1}}
\put (2.5,2){\circle*{.1}}

\thicklines

\put (-2.5,-2) {\line(1,0){1}}\put (-1.5,-1) {\line(1,0){1}}

\put (-.5,0) {\line(1,0){1}}

\put (.5,1) {\line(1,0){1}}\put (1.5,2) {\line(1,0){1}}

\end{picture}

\end{picture}

\vspace*{.5cm}
{\small
\noindent FIG. 2.
The dot
\setlength{\unitlength}{1cm}
\begin{picture}(.2,.2)(0,-.1)
\put (0,0){\circle*{.1}}
\end{picture}
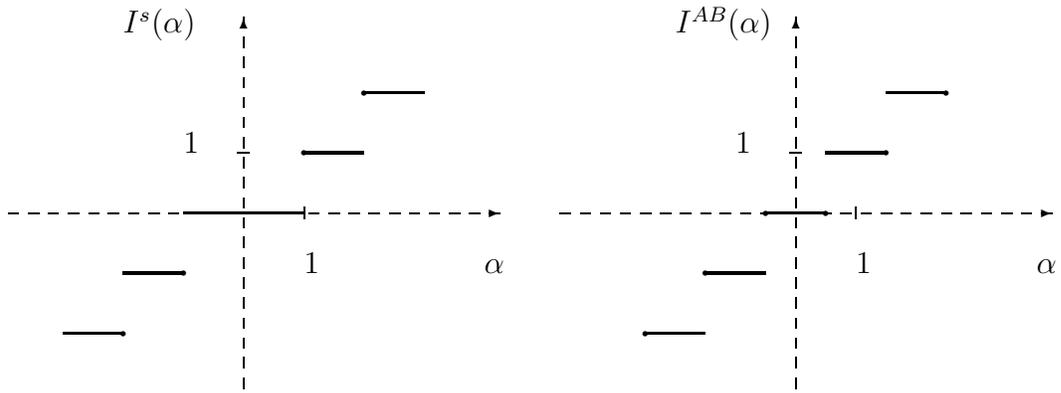
indicates the value of $I(\alpha)$ at the steps.}


\newpage

\begin{thebibliography}{99}

\bibitem{pag}
L.\ Page,
Phys.\ Rev.\ {\bf 36},
444 (1930).

\bibitem{lan} L.\ D.\ Landau, E.\ M.\ Lifshitz:
{\em Quantum Mechanics},
Pergamon Press (1965).

\bibitem{lau}
R.\ B.\ Laughlin,
Phys.\ Rev.\ Lett.\ {\bf 50},
1395 (1983).

\bibitem{qhe1}
R.\ E.\ Prange, S.\ M.\ Girvin (editors):
{\em The Quantum Hall Effect},
Springer (1987).

\bibitem{jai}
J.\ K.\ Jain,
Phys.\ Rev.\ Lett.\ {\bf 63},
199 (1989).

\bibitem{lew}
R.\ R.\ Lewis,
Phys.\ Rev.\ A {\bf 28},
1228 (1983).

\bibitem{aha}
Y.\ Aharonov, D.\ Bohm,
Phys.\ Rev.\ {\bf 115},
485 (1959).

\bibitem{mor}
A.\ Moroz,
Phys.\ Rev.\ A {\bf 53},
669 (1996).

\bibitem{as}
M.\ F.\ Atiyah, I.\ M.\ Singer,
Bull.\ Am.\ Math.\ Soc.\ {\bf 69},
422 (1963).

\bibitem{aps}
M.\ F.\ Atiyah, V.\ K.\ Patodi, I.\ M.\ Singer,
Math.\ Proc.\ Camb.\ Phil.\ Soc.\ {\bf 77},
43 (1975).

\bibitem{cas}
Y.\ Aharonov, A.\ Casher,
Phys.\ Rev.\ A {\bf 19},
2461 (1979).

\bibitem{jac}
R.\ Jackiw,
Phys.\ Rev.\ D {\bf 29},
2375 (1983).

\bibitem{sto}
M.\ Stone,
Ann.\ Phys.\ {\bf 155},
56 (1984).

\bibitem{for}
P.\ Forgacs, L.\ O'Raifeartaigh, A.\ Wipf,
Nucl.\ Phys.\ B {\bf 293},
559 (1987).

\bibitem{gen}
L.\ E.\ Gendenshtein, I.\ V.\ Krive,
Soviet Physics Uspekhi {\bf 28},
645 (1985).

\bibitem{alf}
M.\ Alford, F.\ Wilczek,
Phys.\ Rev.\ Lett.\ {\bf 62},
1071 (1989).

\bibitem{sou}
Ph.\ de Sousa Gerbert,
Phys.\ Rev.\ D {\bf 40},
1346 (1989).

\bibitem{neu}
J.\ von Neumann:
{\em Mathematical foundations of quantum mechanics},
Princeton University Press (1955).

\bibitem{nic}
H.\ Nicolai,
J.\ Phys.\ A {\bf 9},
1497 (1976).

\bibitem{wit}
E.\ Witten,
Nucl.\ Phys.\ B {\bf 185},
513 (1981).

\bibitem{gel}
I.\ M.\ Gel'fand, G.\ E.\ Shilov:
{\em Generalized Functions, Vol.\ I},
 Academic Press (1964).

\bibitem{abr}
M.\ Abramowitz, I.\ A.\ Stegun (editors):
{\em Handbook of Mathematical Functions},
Dover Publications Inc.\ (1965).

\bibitem{sze}
G.\ Szeg\"o:
{\em Orthogonal Polynomials},
Colloquium Publications, American Mathematical Society (1978).

\bibitem{egu}
T.\ Eguchi, P.\ B.\ Gilkey, A.\ J.\ Hanson,
Phys.\ Rep.\ {\bf 66},
213 (1980).

\bibitem{wit2}
E.\ Witten,
Nucl.\ Phys.\ B {\bf 202},
253 (1982).

\bibitem{new1}
R.\ G.\ Newton,
Phys.\ Rev.\ {\bf  96},
523 (1954).

\bibitem{jan}
B.\ Jancovici,
Phys.\ Rev.\ D {\bf 187},
2275 (1969).

\bibitem{sch}
J.\ Schwinger:
{\em Particles, Sources and Fields, Vol.\ III},
Addison-Wesley-Publishing Company (1989).

\end{thebibliography}
\end{document}